\documentclass[10pt, pra,aps,twocolumn,groupedaddress,showpacs]{revtex4-1}

\usepackage{graphicx}
\usepackage{bm}
\usepackage{amssymb,amsfonts,amsmath}
\usepackage[english]{babel}

\begin{document}
\renewcommand{\vec}[1]{\boldsymbol{#1}}
\newcommand{\xx}{\vec{x}}
\newcommand{\rr}{\vec{r}}
\newcommand{\zz}{\vec{z}}
\newcommand{\pp}{\vec{p}}
\newcommand{\qq}{\vec{q}}
\newcommand{\ud}{\mathrm{d}}
\newcommand{\ui}{\mathrm{i}}
\newcommand{\ue}{\mathrm{e}}
\newcommand{\aad}{a/a_d}
\newcommand{\apf}{a_\text{pb}}
\newcommand{\acrit}{a_\text{crit}}
\newcommand{\abif}{a_\text{bif}}
\newcommand{\RD}{\mathcal{R}_D}
\newcommand{\abs}[1]{\left|#1\right|}
\newcommand{\etal}{\emph{et al.}}
\newcommand{\kB}{k_\mathrm{B}}
\newcommand{\Tc}{T_\text{c}}
\newcommand{\Eddagger}{E^\ddagger}
\newcommand{\FIG}{Fig.}
\newcommand{\FIGS}{Figs.}
\newcommand{\SEC}{Sec.}
\newcommand{\EQ}{Eq.}
\newcommand{\EQS}{Eqs.}
\newcommand{\REF}{Ref.}
\newcommand{\REFS}{Refs.}
\newcommand{\p}{\partial}

\newcommand{\ie}{i.\;\!e.}
\newcommand{\eg}{e.\;\!g.}

\title{Transition states and thermal collapse of dipolar Bose-Einstein
condensates}

\author{Andrej Junginger}
\email{andrej.junginger@itp1.uni-stuttgart.de}
\author{Manuel Kreibich}
\author{J\"org Main}
\author{G\"unter Wunner}
\affiliation{Institut f\"{u}r Theoretische Physik 1, 
Universit\"{a}t Stuttgart, 
70550 Stuttgart, Germany}
\date{\today}

\begin{abstract}
We investigate thermally excited, dipolar Bose-Einstein condensates.
Quasi-particle excitations of the atomic cloud cause density fluctuations which
can induce the collapse of the condensate if the inter-particle interaction is
attractive. Within a variational approach, we identify the collectively excited
stationary states of the gas which form transition states on the way to the
BEC's collapse.
We analyze transition states with different $m$-fold rotational symmetry and
identify the one which mediates the collapse. The latter's symmetry depends on
the trap aspect ratio of the external trapping potential which determines the
shape of the BEC. Moreover, we present the collapse dynamics of the BEC and
calculate the corresponding decay rate using transition state theory.
We observe that the thermally induced collapse mechanism is important near the
critical scattering length, where the lifetime of the condensate can be
significantly reduced.
Our results are valid for an arbitrary strength of the dipole-dipole
interaction. Specific applications are discussed for the elements
$^\text{52}$Cr, $^\text{164}$Dy and $^{168}$Er with which dipolar BECs have been
experimentally realized.
\end{abstract}



\maketitle

\section{Introduction}

The Bose-Einstein condensation of atoms with considerable magnetic dipole moment
such as $^\text{52}$Cr, $^\text{164}$Dy and $^{168}$Er atoms
\cite{Griesmaier2005, Lu2011, Aikawa2012} has opened new perspectives in the
field of ultra-cold quantum gases. Due to their electronic structure, these
atoms exhibit magnetic moments of several Bohr magnetons, so that the bosons
interact significantly via the long-range and anisotropic dipole-dipole
interaction (DDI) in addition to the occurrence of scattering processes.
As a consequence of the anisotropic DDI, a wealth of new phenomena emerges in
dipolar Bose-Einstein condensates (BECs). These include isotropic as well as
anisotropic solitons \cite{Pedri2005,Nath2009, Tikhonenkov2008}, biconcave or
structured ground state density distributions \cite{Dutta2007, Ronen2007,
Goral2000}, stability diagrams that crucially depend on the trap geometry
\cite{Koch2008, Santos2000, Goral2002}, radial and angular rotons
\cite{Ronen2007, Santos2003, Wilson2008}, and anisotropic collapse dynamics
\cite{Metz2009, Lahaye2008}.

An important issue in the field of ultra-cold quantum gases is the stability of
the gas. A BEC is a metastable state and several mechanisms can contribute to
its decay, \eg\ inelastic three-body collisions, dipolar relaxation
\cite{Hensler2003}, macroscopic quantum tunneling \cite{Stoof1997,
Marquardt2012}, or the decrease of the s-wave scattering length below its
critical value \cite{Ronen2007}.
Another decay mechanism is the coherent collapse of the atomic cloud due to
thermal excitations. This process is based on the fact that quasi-particle
excitations in an excited BEC lead to time-dependent density fluctuations of the
gas. If the inter-particle interaction is attractive, these fluctuations can
induce the collapse of the condensate, when the density locally becomes high
enough so that the attraction cannot be compensated anymore by the quantum
pressure.
This process is important near the critical scattering length where the
attraction between the bosons becomes dominant. Hints to a thermally induced
collapse can be found in the experiment by Koch \etal\ \cite{Koch2008}. They
measured the value of the critical scattering length and, for a wide range of
the trap aspect ratio, they obtained values which were larger than predicted in
their theoretical investigations.

In a recent publication \cite{Junginger2012d} we have shown qualitatively within
a simple variational approach that the thermally induced collapse of dipolar
BECs is fascinating, because the collapse  dynamics can -- depending on the
external physical parameters -- break the symmetry of the confining trap.
In this paper we present investigations of a thermally excited dipolar BEC
within an \emph{extended} variational approach. Far beyond our previous
investigations using a single Gaussian trial wave function \cite{Junginger2012d}
the extended approach is capable to also reproduce the biconcave shape of the
ground state wave function occurring for certain trapping parameters
\cite{Rau2010b} and to describe complex dynamics of the dipolar BEC. The latter
include the local collapse or collective oscillations and elementary excitations
with arbitrary $m$-fold rotational symmetry \cite{Kreibich2013a}.
Moreover, stability analyses \cite{Rau2010b, Kreibich2013a} within the extended
approach have revealed qualitative differences as compared to the single
Gaussian ansatz: A pitchfork bifurcation of the BEC's ground state is
responsible for the instability of the dipolar BEC below a certain value of the
scattering length and a whole cascade of bifurcations occurs when the scattering
length is further decreased.
The occurrence of these bifurcations gives rise to the expectation that, in
addition to its ground state, further (unstable) stationary states exist in
dipolar BECs and that it is exactly these states which form the transition
states of the condensate on the way to its collapse.

For our investigations, we consider temperatures $T$ in the region
$0 < T_0 < T \ll \Tc$.
Below a temperature $T_0$, collective oscillations of the BEC are not
thermally excited and macroscopic quantum tunneling will be the dominant decay
mechanism. At $T>T_0$, the collective dynamics of the condensate is excited, and
above $\Tc$ the BEC does not exist anymore.
At temperatures $T \ll \Tc$ small compared to the critical one, the
condensate is almost pure and
excitations of single bosons to higher quantum states can be neglected, so that
the condensate can be described in a mean-field approximation by the
Gross-Pitaevskii equation (GPE).
At temperatures which are on the order of the critical temperature, $T \approx
\Tc$, a significant number of bosons will occupy excited states and
Hartree-Fock-Bogoliubov theory \cite{Proukakis1996,Griffin1996} can be applied.
Also c-field methods \cite{Blakie2008} allow for the investigation of the
finite-temperature BEC up to the critical temperature when the contribution of
incoherent particles in the gas is of importance.
This regime is, however, not subject of our investigations.

Our paper is organized as follows:
In \SEC\ \ref{sec:theory}, we present the theoretical description of the dipolar
BEC within the framework of the GPE and introduce particle number scaled units.
Moreover, the variational approach to the condensate's wave function as well as
the calculation of the thermal decay rate by applying transition state theory
(TST) is demonstrated.
The results are presented in \SEC\ \ref{sec:results}. Here we discuss the
presence of various transition states with different $m$-fold rotational
symmetry, the dynamics of BECs which are excited above the corresponding
activation energy and the decay rate of the condensate.
Our results are valid for an arbitrary strength of the dipole-dipole interaction
and, at the end of \SEC\ \ref{sec:results}, we discuss applications to the
elements $^{52}$Cr, $^\text{164}$Dy and $^{168}$Er which possess different
magnetic moments.

\section{Theory}\label{sec:theory}

\subsection{The dipolar BEC} \label{sec:dipolar-BEC}

If the temperature of the quantum gas is small compared to the critical
temperature $\Tc$, then excitations of single bosons to higher quantum states
can be neglected and the atomic cloud can be described by a single order
parameter $\psi(\rr,t)$ which is a solution of the time-dependent GPE
\begin{equation}%
\begin{split}
\ui \partial_{t} \psi(\rr,t)
&=
\hat H \psi(\rr,t) \\
&=
\left( - \Delta + V_\text{ext} + V_\text{int}\right) \psi(\rr,t) \, .
\label{eq:GPE}
\end{split}
\end{equation}
Here, the terms $V_\text{ext}$ and $V_\text{int}$ describe the interactions of
the bosons with an external trapping potential as well as the inter-particle
interaction,
\begin{align}%
V_\text{ext} =&\,
\gamma_\rho^2 \rho^2 \Big[ 1 + s \cos(m\phi) \Big]
+ \gamma_z^2 z^2 \, ,
\label{eq:Vext}\\
\begin{split}
V_\text{int} =&\,
8 \pi a \abs{\psi(\rr,t)}^2 \\
& + \int \! \ud^3 \rr'\, \frac{1-3\cos^2
\theta}{\abs{\rr-\rr'}^3} \,
\abs{\psi(\rr',t)}^2 \, .
\label{eq:Vint}
\end{split}
\end{align}
In \EQ\ \eqref{eq:Vext}, $\gamma_{\rho,z}$ denote the strength of the external
trapping potential which is harmonic in both the radial and $z$-direction and
cylindrically symmetric for $s=0$. For $s \neq 0$, this symmetry can be broken
in a way that the trap exhibits an $m$-fold rotational symmetry. However, note
that we always investigate condensates in axisymmetric traps ($s=0$) in this
paper and that the additional term for $s \neq 0$ is only used to specifically
access the excited states as discussed in the appendix.
The inter-particle interaction $V_\text{int}$ on the one hand takes into account
low-energy scattering processes between two bosons which are described by the
s-wave scattering length $a$. On the other hand, it describes the long-range DDI
between the bosons which we assume to be aligned along the $z$-direction by an
external magnetic field ($\theta$ is the angle between the $z$-axis and the
vector $\rr - \rr'$).

The GPE \eqref{eq:GPE} and the interaction terms
\eqref{eq:Vext}--\eqref{eq:Vint} are given in dimensionless units which we
obtained by measuring lengths in terms of the ``dipole length'' $a_d = m \mu_0
\mu^2/(2\pi\hbar^2)$, energies in units of $E_d=\hbar^2 / (2 m a_d^2)$,
frequencies in terms of $\omega_d = E_d/\hbar$ and inverse temperatures in
multiples of $\beta_d=E_d^{-1}$.
Furthermore, we applied a particle number scaling
\begin{equation}
  (\rr, \psi, E, \beta, \omega) \longrightarrow
  (N \rr, N^{-3/2} \psi, N^{-1} E, N \beta, N^{-2}\omega)
\end{equation}
in order to eliminate the explicit occurrence of the particle number $N$ in the
interaction terms and define the trapping parameters as $\gamma_{\rho,z} =
\omega_{\rho,z} / (2N^2 \omega_d)$ where $\omega_{\rho,z}$ are the trap
frequencies in SI units.

\subsection{Variational approach to the GPE} \label{sec:var-approach}

\subsubsection{The time-dependent variational principle}

There are several ways to solve the GPE \eqref{eq:GPE}, one of which is the
discretization of the wave function on grids. The dynamics can then be
calculated using the split-operator method and the ground state of the BEC is
accessible via an imaginary time evolution of \EQ\ \eqref{eq:GPE}.

For our purpose, a more powerful approach is the solution of the
GPE within a variational framework, because this allows to access collectively
excited stationary states of the GPE which play a crucial role in our
investigation.
In this variational approach, the time-dependent wave function $\psi(\rr,t) =
\psi(\rr, \zz(t))$ describing the BEC is expressed in terms of a set of
time-dependent and complex variational parameters $\zz(t)$. An approximate
solution of the GPE in the Hilbert subspace spanned by the variational
parameters is then given by the McLachlan variational principle
\cite{McLachlan1964}. This claims the norm of the difference between the left-
and the right-hand side of the GPE to be minimal, \ie\
\begin{equation}
   \| \ui \dot \psi - \hat H \psi \| \stackrel{!}{=} \mathrm{min},
   \label{eq:McLachlan-variational-principle}
\end{equation}
where the variation is carried out with respect to $\dot\psi$ (for simplicity,
we have omitted the arguments $(\rr,t)$ of the wave function). Applying this
variational principle to a parameterized wave function $\psi(\rr,\zz(t))$, one
obtains the set of ordinary first order differential equations
\cite{Cartarius2008a}
\begin{equation}
\left< \left.
   \frac{\p \psi}{\p \zz}
\right|
   \frac{\p \psi}{\p \zz}
\right>
\dot \zz = - \ui
\left< \left. \left.
   \frac{\p \psi}{\p \zz}
\right| \hat H \right|
   \psi
\right> 
\label{eq:equations-of-motion}
\end{equation}
which determine the time evolution of the wave function. The mean-field energy
of the condensate for a given set of variational parameters is obtained from
the energy functional
\begin{equation}
   E(\zz) =
   \Bigl< \psi(\rr,\zz(t)) \Bigl|
      -\Delta
      +V_\text{ext}
      + \frac{1}{2} V_\text{int}
   \Bigr| \psi(\rr,\zz(t)) \Bigr> \, .
   \label{eq:emf}
\end{equation}

\subsubsection{The variational ansatz}

The dipolar BEC, described by the GPE \eqref{eq:GPE} with the interaction
potentials \eqref{eq:Vext}--\eqref{eq:Vint} and $s=0$, naturally exhibits a
cylindrical symmetry due to the alignment of the dipoles along the $z$-direction
and due to the shape of the external trapping potential. In this case, the
$z$-component of the angular momentum with eigenfunctions $\exp(\ui m \phi)$ and
the corresponding quantum number $m$ are conserved. A suitable choice for the
trial wave function is, therefore, given by the coupled Gaussians
\cite{Kreibich2013a}
\begin{equation}
\begin{split}
\psi(\rr,\zz(t))
=
\sum_{i=1}^{N_g} &
\biggl(1 + \sum_{m\neq0} \sum_{p=0,1}
d^i_{m,p} \rho^{\abs{m}} z^p \ue^{\ui m \phi} \biggr) \\
& \times \exp \Bigl( A_\rho^i \rho^2 + A_z^i z^2 + p_z^i z + \gamma^i \Bigl)
\label{eq:ansatz-arbitrary-m}
\end{split}
\end{equation}
with variational parameters $(d^i_{m,p}, A_\rho^i, A_z^i, p_z^i, \gamma^i)$.
Here, the $A_\rho^i, A_z^i$ describe the Gaussians' width in $\rho$- and
$z$-direction, $p_z^i$ the displacement of the BEC along the $z$-axis
and the $\gamma^i$ give the amplitude and phase of each Gaussian. Furthermore,
the $d^i_{m,p}$ describe the non-axisymmetric contributions to the wave
function with arbitrary angular momentum $m$ and even ($p=0$) or odd ($p=1$)
parity.
We note that the ansatz
\begin{equation}
\psi(\rr,\zz(t))
=
\sum_{i=1}^{N_g}
\exp\left(A_x^i x^2 + A_y^i y^2 + A_z^i z^2 + \gamma^i\right)
\label{eq:ansatz-kartesian}
\end{equation}
with variational parameters $(A_x^i, A_y^i, A_z^i, \gamma^i)$ can be used as an
alternative to \EQ\ \eqref{eq:ansatz-arbitrary-m}, if only excitations with
$m=0$ ($A_x^i=A_y^i$) and $m=2$ ($A_x^i \neq A_y^i$) rotational symmetry are of
interest (see below).

Both the trial wave functions \eqref{eq:ansatz-arbitrary-m} and
\eqref{eq:ansatz-kartesian} have the advantage that all the integrals occurring
in the equations of motion \eqref{eq:equations-of-motion} and the mean-field
energy \eqref{eq:emf} can be evaluated (semi-)analytically. For their
calculation, we refer the reader to \REFS\ \cite{Rau2010a, Kreibich2013a} where
they have been used in previous investigations.

\subsection{Thermal decay rate}

As we will discuss in \SEC\ \ref{sec:results}, the GPE \eqref{eq:GPE} possesses
-- depending on the external physical parameters -- one ground state and several
excited stationary states with different $m$-fold rotational symmetry.
While the ground state is stable with respect to all local variations of the
variational parameters, the excited states possess an unstable direction.
Therefore the latter form saddle points of the energy functional \eqref{eq:emf}
which have the physical meaning of ``energy barriers''.
If the excitation energy of the BEC is larger than this barrier, it can be
crossed along the unstable direction which results in the coherent collapse of
the condensate as we will discuss below.
In terms of reaction dynamics, each of the collectively excited states,
therefore, forms a transition state (TS) on the way to the collapse of an
excited BEC.

Having identified the TS, the corresponding reaction rate can be calculated by
applying TST. This is possible after having performed a change of variables by
means of a normal form expansion \cite{Junginger2012a, Junginger2012b} which
maps the variational parameters to local canonical variables $(\pp,\qq)$.
In these the calculation of the flux over the saddle at an excitation energy
larger than the saddle point energy is straightforward. Considering a
condensate at an inverse temperature of $\beta = 1/\kB T$, the respective
reaction rates at different excitation energies have to be Boltzmann
averaged, and within a harmonic approximation of the ground state and the
saddle one obtains the result \cite{Haenggi1990}
\begin{equation}
\Gamma = \frac{1}{2\pi} \,
\frac{\prod_{i=1}^d \omega_i}
     {\prod_{i=2}^d \omega'_i}
\,
\ue^{- \beta E^\ddagger} \,.
\label{eq-Gamma-conventional}
\end{equation}
Here, the $\omega_i$ denote the stable oscillation frequencies in the vicinity
of the ground state, $\omega'_i$ those at the excited state, and $E^\ddagger$ is
the height of the energy barrier.

\section{Results}\label{sec:results}

In this section, we present the results of our investigations of thermally
excited dipolar BECs. We give a detailed discussion of the collectively excited
states, point out the one which mediates the collapse, and calculate the
corresponding collapse dynamics. We investigate the behavior of the energy
barrier, and calculate the thermal decay rate for experimentally relevant
temperatures.

Our main focus will be on the dependence of the different properties on the trap
aspect ratio $\lambda = \gamma_z/\gamma_\rho$ and we therefore parameterize
the external trapping parameters $\gamma_{\rho,z}$ by their mean value
$\bar\gamma = (\gamma_\rho^2 \gamma_z)^{1/3}$.
Our focus on the trap aspect ratio is because dipolar BECs exhibit the
surprising feature that, for certain values of $\lambda$, they exhibit a
blood-cell shaped ground state density distribution \cite{Ronen2007, Rau2010b}.
In contrast to ``conventional'' density distributions where the maximum density
of the atomic cloud is located in the center of the trap, such blood-cell shaped
BECs show their maximum density away from the center.

We investigate both the regime with a conventional density distribution and the
blood-cell-like regime, and as exemplary values for these regimes, we choose the
trap aspect ratios $3 \leq \lambda \leq 4.5$ and $7 \leq \lambda \leq 8$. If
not stated otherwise, the calculations presented in this section are performed
using a number of $N_g = 6$ coupled Gaussians which have proven their capability
to reproduce or even to exceed the results of numerical grid calculations
\cite{Rau2010b}.

\subsection{Transition states of dipolar BECs}\label{sec:stat-states}

\begin{figure}[t]
\includegraphics[width=\columnwidth]{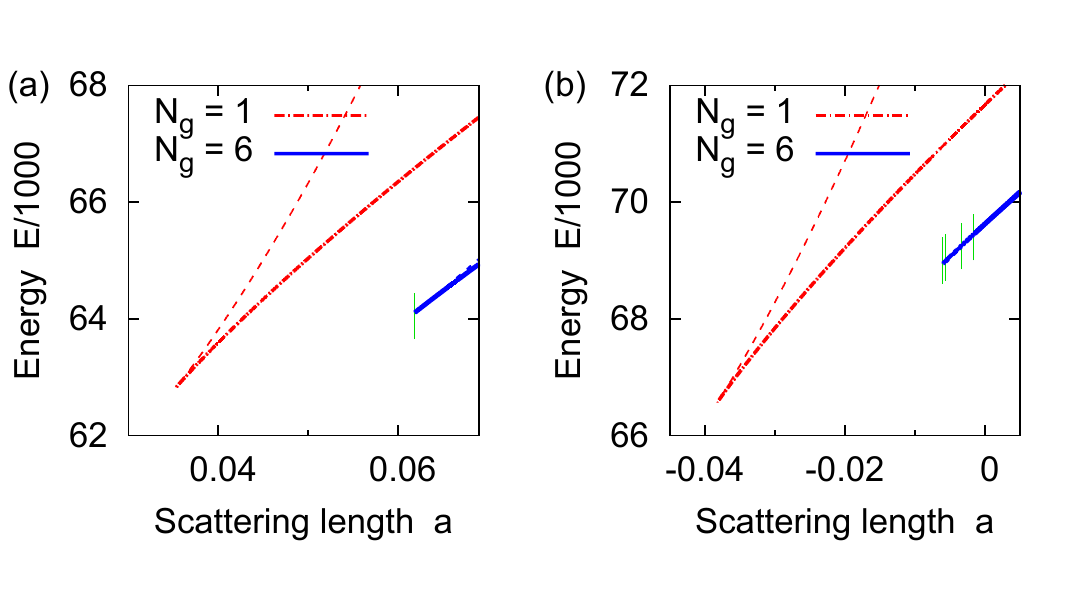}
\caption{%
(Color online) 
Mean field energy $E$ of the stationary states of a dipolar BEC for a
mean trap frequency $\bar\gamma = 8000$ and a trap aspect ratio of
$\lambda=4$ (a) and $\lambda=7$ (b). Shown are the values obtained from the
simple variational ansatz using a single Gaussian wave function ($N_g=1$) and
the extended approach using the trial wave functions in \EQS\
\eqref{eq:ansatz-arbitrary-m}--\eqref{eq:ansatz-kartesian} ($N_g=6$).
The solid and dashed-dotted lines represent the ground state of the condensate
and the dashed lines the collectively excited states.
Using the simple variational ansatz, the energy difference between the ground
and the excited state increases very rapidly while it is smaller than the
linewidth of the plot within the extended approach.
The vertical lines indicate the positions of the bifurcations (cf.\ \FIG\
\ref{fig:m-modes}).
}
\label{fig:comparison}
\end{figure}

As we have mentioned above, the GPE possesses, in general, several stationary
states which are fixed points ($\dot\zz = 0$) of the equations of motion
\eqref{eq:equations-of-motion}. The existence of these states depends crucially
on the physical parameters of the system, namely the external trapping
parameters as well as the s-wave scattering length $a$ which can be varied by
means of Feshbach resonances in an experiment.
Below a critical value $\acrit$ the BEC cannot exist: From a physical point of
view, the reason is that the attractive inter-particle interaction becomes so
strong that it cannot be compensated anymore by the quantum pressure.
Mathematically speaking, this is because the stable ground state of the
condensate undergoes a bifurcation, in which it either becomes unstable or it
vanishes completely.

Figure \ref{fig:comparison} shows the stationary states of the dipolar BEC
obtained from a single ($N_g=1$) Gaussian trial wave function and
$N_g=6$ coupled wave functions in \EQS\
\eqref{eq:ansatz-arbitrary-m}--\eqref{eq:ansatz-kartesian}. We choose a mean
trap strength of $\bar \gamma = 8000$ and a trap aspect ratio
of $\lambda=4$ (conventional density distribution) as well as $\lambda=7$
(blood-cell shaped BECs).
Using a single Gaussian wave function (dashed-dotted and dashed lines), the
ground state and an excited state emerge in a tangent bifurcation at a
scattering length of $a \approx 0.0353$ ($\lambda=4$) and $a \approx -0.0382$
($\lambda=7$), respectively. The energy difference between the two states
increases rapidly when the scattering length is increased.

Within the extended variational approach (solid line), both the value of the
mean field energy and the position of the tangent bifurcation are shifted.
Moreover, Rau \etal\ \cite{Rau2010b} and Kreibich \etal\ \cite{Kreibich2013a}
showed that, for BECs with a biconcave shape, there is a whole cascade of
stability changes of the condensate's ground state with respect to excitations
with different $m$-fold rotational symmetry (the positions are indicated by the
vertical bars in \FIG\ \ref{fig:comparison}).
Each of these is accompanied by with a bifurcation, in which two additional
excited but unstable stationary states emerge. An important property of these
states which have not yet been discussed in the literature is the fact that the
energy difference with respect to the ground state is several orders of
magnitude smaller than the one obtained from the simple variational ansatz.
As we have already mentioned above, these excited states form transition states
of the condensate's collapse dynamics, and because of the smaller energy barrier
the corresponding reaction rates will be significantly higher than our
calculations within the single Gaussian trial wave function
\cite{Junginger2012d} have predicted.

\begin{figure}[t]
\includegraphics[width=\columnwidth]{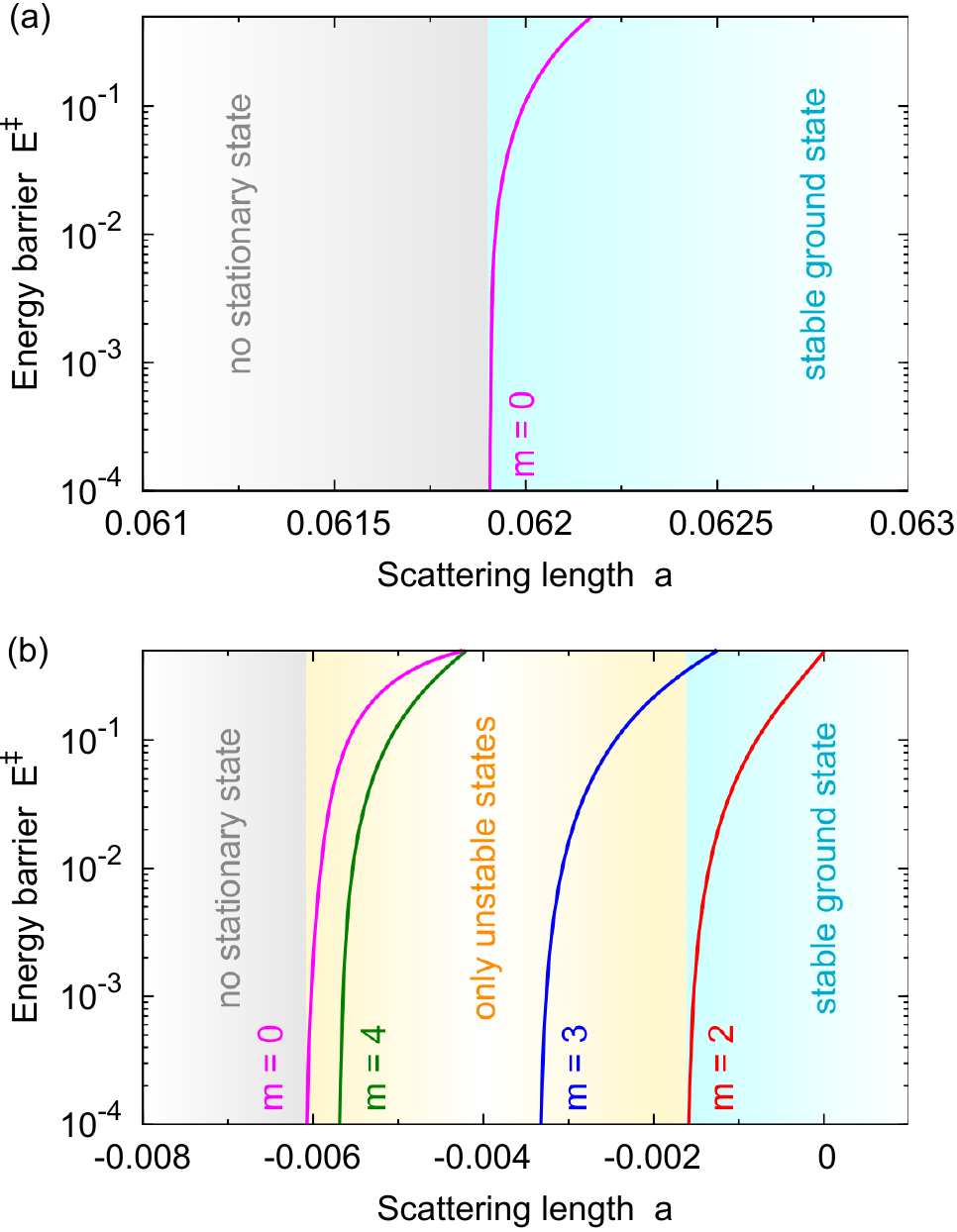}
\caption{%
(Color online) 
Energy barrier $E^\ddagger = E_\text{ex} - E_\text{gs}$ of a dipolar BEC for a
mean trap frequency $\bar\gamma = 8000$ and a trap aspect ratio of
$\lambda=4$ (a) and $\lambda=7$ (b). Both sub-figures show the region of the
s-wave scattering length near the critical scattering length $\acrit$.
For $\lambda=4$ and a scattering length above the critical value ($\acrit
\approx 0.0619$) there exist only the ground state of the BEC and a
collectively excited state with $m=0$ symmetry.
For $\lambda=7$ several collectively excited states with different symmetry
$m=0,2,3,4$ are present. The one which is involved in the bifurcation at the
critical scattering length ($\acrit \approx -0.0016$) is the
$m=2$
excited state and it is this state which has the smallest energy barrier in the
parameter region where the BEC is stable. (See text for further explanations.)
}
\label{fig:m-modes}
\end{figure}

The essential property of the dipolar BEC concerning its thermally induced
collapse is the height of the energy barrier $E^\ddagger = E_\text{ex} -
E_\text{gs}$ because the reaction rate \eqref{eq-Gamma-conventional} depends
exponentially on this quantity. Therefore, in the following we discuss the
energy barriers in more detail.
\FIG\ \ref{fig:m-modes}(a) shows the situation at a trap aspect ratio of
$\lambda = 4$. Here the ground state of the BEC and an excited state which also
has an axial symmetry ($m=0$) emerge in a tangent bifurcation at a scattering
length of $\acrit \approx 0.0619$ and the energy barrier
$E^\ddagger$ between these states increases quickly by several orders of
magnitude when the scattering length $a$ is increased.
Below the critical scattering length, there exists no state and above this value
the $T=0$ ground state is stable over the whole range of the scattering length.
Other excited states with $m \neq 0$ do not participate in bifurcations together
with the ground state.

The situation is different when we consider blood-cell shaped BECs at a trap
aspect ratio of $\lambda = 7$ [see \FIG\ \ref{fig:m-modes}(b)]. Decreasing the
scattering length from the region where a stable BEC exists the ground state
becomes unstable with respect to elementary excitations with $m=2$ rotational
symmetry at a critical scattering length of $\acrit \approx -0.0016$. At the
same point, an excited state bifurcates from the ground state which also exists
at values $a > \acrit$.
In contrast to the case $\lambda =4$ where the ground state vanishes in
a tangent bifurcation as discussed above, the ground state changes its
stability in a pitchfork bifurcation at $\lambda = 7$ and persists
as a stationary but unstable state also below this critical value.
Due to the instability of all the states at $a < \acrit$, this region is, of
course, not relevant for experiments. However, it is relevant from a theoretical
point of view, because other excited states which exist in the stable regime $a
> \acrit$ can emerge there in further bifurcations.
For the blood-cell shaped BEC in \FIG\ \ref{fig:m-modes}(b), three more
bifurcations occur: At a scattering length of $a \approx -0.0033$ two excited
states with $m=3$ rotational symmetry bifurcate in a pitchfork bifurcation.
Moreover, we observe the pitchfork bifurcation of two $m=4$ states at $a \approx
-0.0057$ and finally the tangent bifurcation together with an $m=0$ excited
state at $ a \approx -0.0061$ below which no stationary state is present
anymore.
Note that all the energy barriers formed by the different excited states
increase quickly by several orders of magnitude if the scattering length is
increased from the respective bifurcation, and that it is the $m=2$ state which
corresponds to the smallest energy barrier.

\subsection{Dynamics of excited BECs} \label{sec:dynamics}

The collectively excited states of the condensate discussed in the previous
section correspond to saddle points of the energy functional \eqref{eq:emf} in
the space of variational parameters, while the condensate's ground state
corresponds to a local minimum. Because energy is conserved, we can discuss the
physical meaning of the excited states in terms of the energy functional and
with it interpret the possible dynamics.
We first consider a BEC in its ground state. In this case, the dynamics
resides at the minimum of the energy functional, and for slightly excited
states the dynamics of the condensate is restricted to a small vicinity of the
energy functional's minimum.
A qualitative change in the dynamics of the BEC, however, occurs when the
excitation energy of the condensate becomes larger than the energy of the
lowest saddle in $\zz$-space. In this case, a new region in the variational
space becomes accessible and transitions between these regions correspond to
crossings of the saddle point.

\begin{figure*}[t]
\includegraphics[width=2\columnwidth]{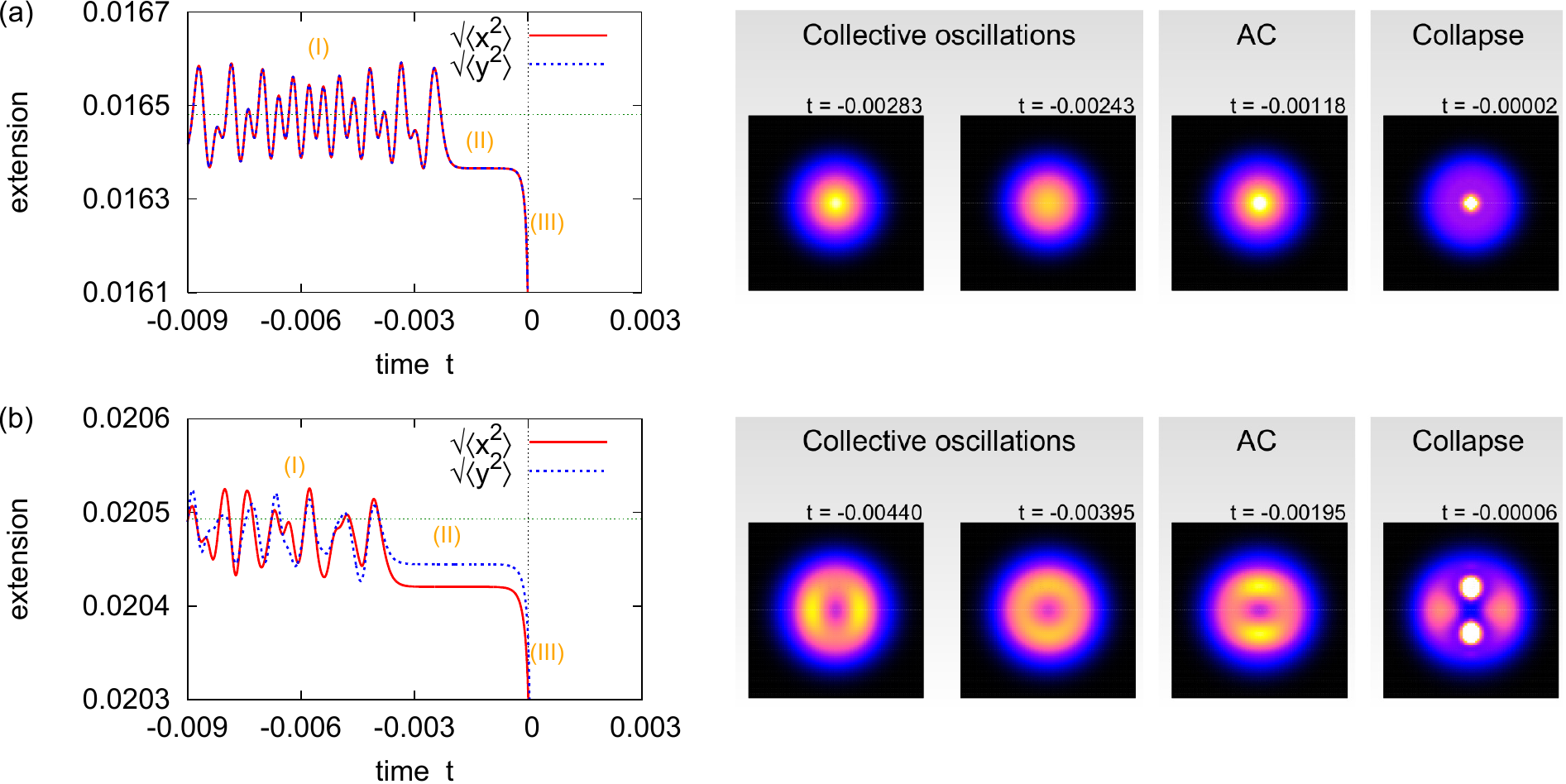}
\caption{%
(Color online) 
Collapse dynamics of an excited dipolar BEC at (a) a trap aspect ratio
$\lambda=4$ and a scattering length of $\aad = 6.233\times 10^{-2}$ as well
as (b) $\lambda=7$ and $\aad = 3.940\times 10^{-4}$.
Shown is the time
evolution of the BEC's rms-extensions $\sqrt{\left< x^2\right>}$ and
$\sqrt{\left< y^2\right>}$, respectively, and the excitation energy is
slightly above the energy barrier which is $E^\ddagger = 1.0$ for both values of
the scattering length given above.
The dynamics shows (I) collective oscillations of the atomic cloud, (II) the
formation of the quasi-stationary activated complex (AC) and (III) the collapse
of the BEC.
}
\label{fig:dynamics}
\end{figure*}

Because the physical meaning of the crossing of the saddle point is not clear
a priori, we present the corresponding dynamics of the BEC in the
following and consider a condensate which is excited to an energy slightly
higher than the lowest saddle point energy. Figure \ref{fig:dynamics} shows the
corresponding dynamics for trap aspect ratios $\lambda=4$ (a) and $\lambda=7$
(b) in terms of the extensions $\sqrt{\left<x^2\right>}$ and
$\sqrt{\left<y^2\right>}$ of the condensate (left-hand side) as well as in terms
of the density profiles of the atomic cloud (right-hand side).

In the case of an excited BEC with conventional ground state density
distribution at $\lambda=4$ [see \FIG\ \ref{fig:dynamics}(a)] we observe at
first collective oscillations of all the atoms (I). When the dynamics reaches
the vicinity of the saddle, the motion turns into a quasi-stationary state (II)
where the extension remains nearly constant for a period of time which is on the
order of a few oscillation periods. At the end, the extension begins to shrink
very rapidly (III) and it further contracts to zero extension (not shown) for $t
\to 0$ which means the collapse of the BEC. Note that the
dynamics is axisymmetric all the time ($\sqrt{\left<x^2\right>} =
\sqrt{\left<y^2\right>}$), so that it is the breathing mode of the BEC which is
associated with the crossing of the saddle, and the subsequent collapse. This
behavior is to be expected since the only TS in this case has an $m=0$
rotational symmetry [cf.\ \FIG\ \ref{fig:m-modes}(a)].
From the density profiles of the atomic cloud [see right-hand side of \FIG\
\ref{fig:dynamics}(a)], we see that the collective oscillations are associated
with small density fluctuations of the BEC. At a certain time ($t = -0.00118$)
the BEC reaches the ``activated complex'' with a critical density in its center
that leads to the dominance of the attractive inter-particle interaction and to
the subsequent collapse for $t \to 0$.

For a trap aspect ratio of $\lambda=7$ [see \FIG\ \ref{fig:dynamics}(b)], we
observe similar dynamics, also consisting of collective oscillations, crossing
the saddle, and the collapse of the BEC.
However, significant differences exist for blood-cell shaped BECs:
At first, the dynamics is not axisymmetric which corresponds to the fact that
the energetically lowest saddle has an $m=2$ rotational symmetry [cf.\ \FIG\
\ref{fig:m-modes}(b)]. The oscillation mode which is responsible for the
collapse is, therefore, the quadrupole mode.
This behavior is astonishing, because the external trapping potential is
axisymmetric, but it confirms hints for a symmetry breaking collapse scenario
that we have obtained within a simple model in a previous investigation
\cite{Junginger2012d}.
Secondly, we observe richer collapse dynamics as can be seen in the density
profiles: The collective oscillations correspond to the dynamics where
local density maxima occur on a ring around the center of the trap. Also the
activated complex (reached at $t = -0.00195$) of the system is a density
distribution which shows two local maxima on this ring. Precisely at these
positions, the attractive interaction becomes dominant and the collapse of the
BEC is now induced locally.
Similar collapse dynamics also occur if one of the other saddles is
crossed (not shown). According to the different rotational symmetry of the
respective transition state, the whole collapse dynamics differs with respect to
this point, meaning that a different number of angular patterns can be
observed. However, independent of the respective rotational symmetry all
these cases result in the collapse of the condensate, so that each of the
states discussed above forms a barrier on the way to the BEC's collapse.

The collapse scenario shown in \FIG\ \ref{fig:dynamics}(b) is similar to the
d-wave collapse investigated by Metz \etal\ \cite{Metz2009}, and also Wilson
\etal\ \cite{Wilson2009} have investigated an angular collapse of dipolar BECs.
We emphasize, however, that the physical situation in which this collapse is
observed is totally different: In \REFS\ \cite{Metz2009, Wilson2009} the stable
$T=0$ ground state of the BEC is considered and the collapse dynamics is
investigated after having ramped down the s-wave scattering length below the
critical value ($a < \acrit$), \ie\ into a region where the condensate cannot
exist anymore.
By contrast, we observe similar dynamics in a region of the physical parameters
where the BEC's ground state is stable, and there is also the quantitative
difference that the collapse dynamics shown in \FIG\ \ref{fig:dynamics} exhibit
an $m=2$ symmetry, whereas Wilson \etal\ observe an $m=3$ collapse in their
different experimental setup. The reason for the collapse of the BEC is in our
case not a change of the physical parameters but the excitation of the
condensate's internal degrees of freedom, namely its collective modes.

\subsection{Activation energy and thermal decay rate}\label{sec:dE-and-rates}

As discussed in \SEC\ \ref{sec:stat-states}, several TSs can exist which lead to
different collapse scenarios with different rotational symmetry. To identify the
one which is important for a BEC in an experiment we have to take into account
the fact that a BEC in an experiment has a finite temperature $T>0$. In contrast
to the bosons in the quantum gas, the quasi-particles of the collective dynamics
are not condensed and, therefore, all collective modes are populated according
to the Boltzmann factor, and the corresponding reaction rates for each mode
depend exponentially on the height of the respective energy barrier $E^\ddagger$
[cf.\ \EQ\ \eqref{eq-Gamma-conventional}].
Therefore, the mode which determines the decay rate of the condensate is always
the one which corresponds to the saddle with the least energy because
contributions from higher saddles are exponentially damped.
At a trap aspect ratio of $\lambda=4$ [\FIG\ \ref{fig:m-modes}(a)] this is the
$m=0$ excited state and for $\lambda=7$ [\FIG\ \ref{fig:m-modes}(b)] it is the
$m=2$ state. The other excited states which exist for $\lambda=7$ form barriers
which lie several orders of magnitude higher, so that they can be neglected.

\begin{figure}[t]
\includegraphics[width=\columnwidth]{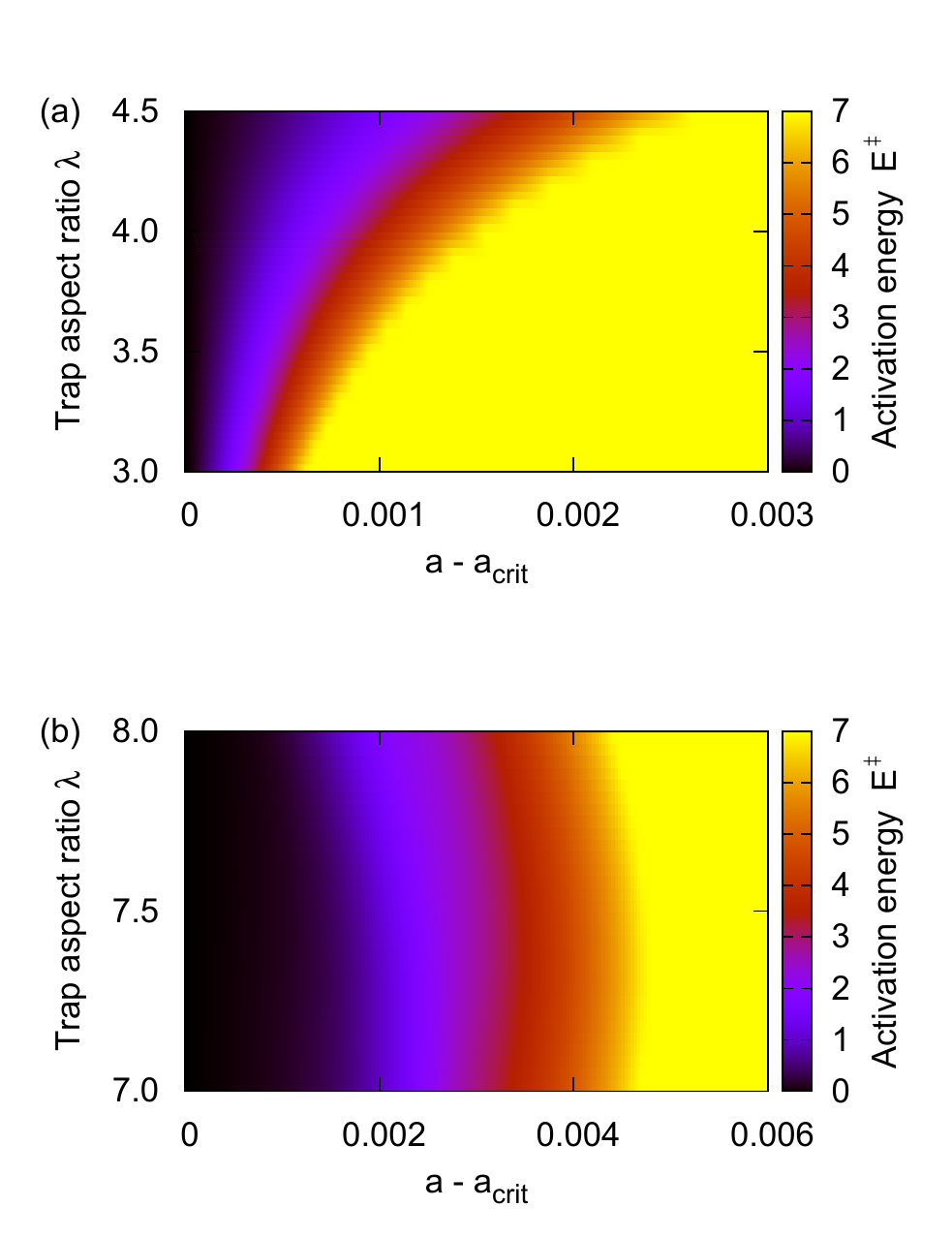}%
\caption{%
(Color online) 
Activation energy $E^\ddagger$ for the thermally induced coherent collapse for
(a) non-blood-cell shaped BECs at trap aspect ratios $3 \leq \lambda \leq 4.5$
and (b) blood-cell shaped BECs at $7 \leq \lambda \leq 8$.
In (a) the energy barrier with increasing scattering length becomes smaller the
higher the trap aspect ratio is.
For blood-cell shaped BECs (b) the behavior of the energy barrier with
increasing scattering length changes only marginally when varying the trap
aspect ratio.
We note that, for a $^{52}$Cr condensate consisting of $N=10\,000$ bosons, the
range of the energy barrier shown ($0 \leq E^\ddagger \leq 7$) corresponds to a
thermal energy $T = E^\ddagger / \kB$ of $0 \leq T \leq 141\,$nK which is about
20\% of the critical temperature ($\Tc = 700\,$nK \cite{Griesmaier2005}). The
range of the scattering length is $0.275$ Bohr radii above the critical
scattering length in (a) and $0.55$ Bohr radii in (b).
}
\label{fig:energy-barrier}
\end{figure}

Figure \ref{fig:energy-barrier} shows the behavior of the activation energy for
two ranges of the trap aspect ratio $3 \leq \lambda \leq 4.5$ and $7 \leq
\lambda \leq 8$.
For a dipolar BEC with conventional density distribution we see in \FIG\
\ref{fig:energy-barrier}(a) that the behavior of the energy barrier
significantly depends on the trap aspect ratio. For smaller trap aspect
ratios $\lambda=3$, we observe a rapid increase of the energy barrier when the
scattering length is increased from the critical value and the energy barrier
reaches a value of $E^\ddagger \approx 7$ already at $a - \acrit = 0.0005$.
At a larger trap aspect ratio $\lambda=4.5$ the energy barrier increases much
slower and reaches the same value only at a scattering length $a - \acrit =
0.0025$, so that the relevant region is about five times larger.
When we consider blood-cell shaped BECs [see \FIG\ \ref{fig:energy-barrier}(b)],
there is only a minor change of the energy barrier's behavior when the trap
aspect ratio is varied. However, the increase of the scattering length is even
slower as compared to \FIG\ \ref{fig:energy-barrier}(a) and we reach a barrier
height of $E^\ddagger \approx 7$ at a scattering length $a - \acrit = 0.0045$
which is about nine times larger than at a trap aspect ratio of $\lambda=3$.

\begin{figure}[t]
\includegraphics[width=\columnwidth]{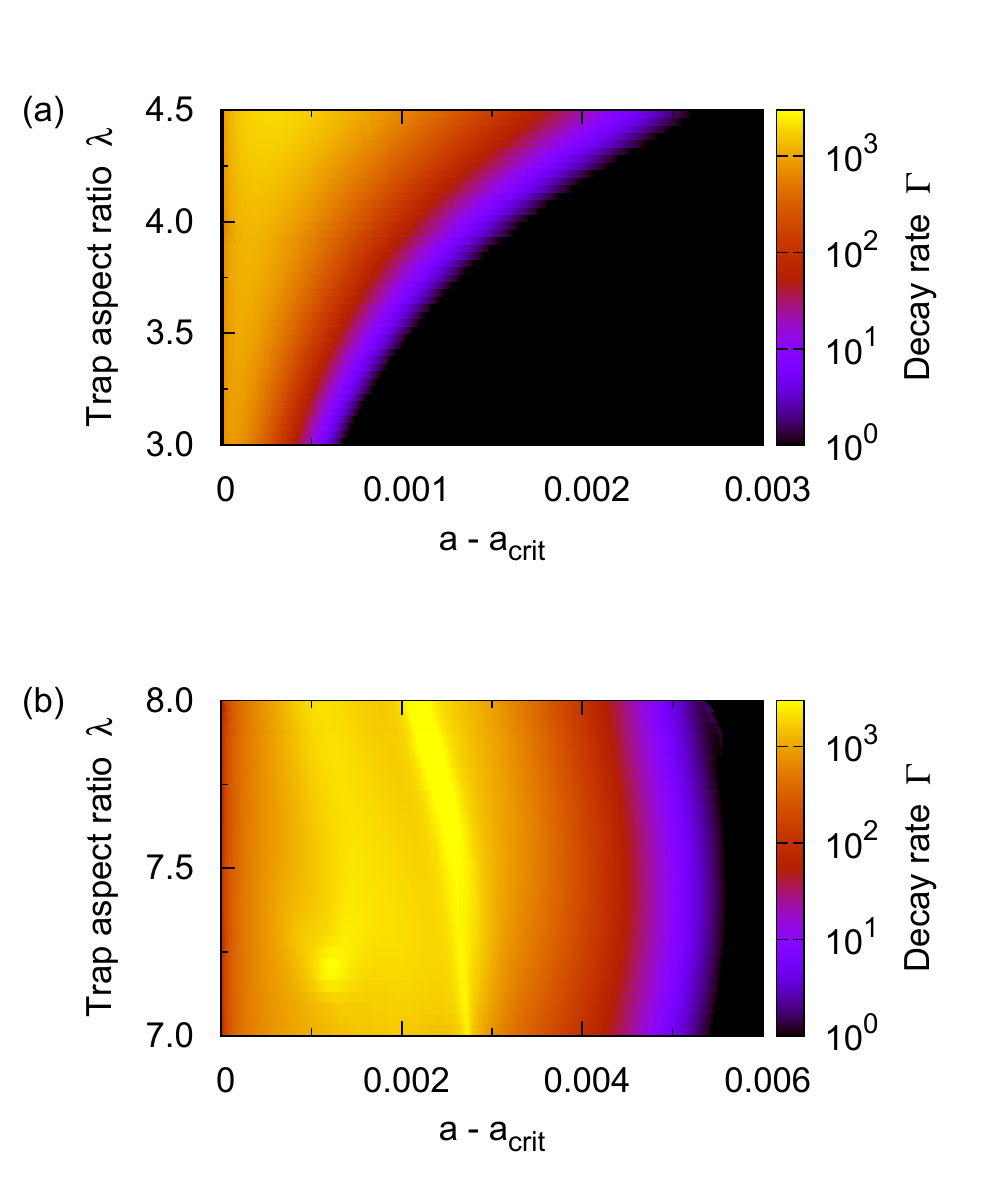}
\caption{%
(Color online) 
Decay rate due to the thermally induced collapse for (a) non-blood-cell shaped
BECs at $3 \leq \lambda \leq 4.5$ and (b) blood-cell shaped BECs at $7
\leq \lambda \leq 8$ at an inverse temperature of $\beta = 1$.
Depending on the trap aspect ratio, the decay rate shows a similar behavior as
the activation energy in \FIG\ \ref{fig:energy-barrier}.
The ranges of the trap aspect ratios and the scattering length are the same as
in \FIG\ \ref{fig:energy-barrier} and the range of the decay rate ($1 \leq
\Gamma \leq 3\times 10^3$) corresponds to mean lifetimes $\tau = \Gamma^{-1}$ of
$1.3\,\text{ms} \leq \tau \leq 3800\,\text{ms}$ for a $^{52}$Cr BEC of
$N=10\,000$ atoms. The inverse temperature $\beta=1$ corresponds to $T=20\,$nK.
}
\label{fig:decay-rate}
\end{figure}

This behavior of the energy barrier is of course directly reflected in the
corresponding thermal decay rate of the dipolar BEC, shown in \FIG\
\ref{fig:decay-rate} for an inverse temperature of $\beta = 1$.
We find the general behavior that there are very small decay rates $\Gamma
\lesssim 1$ far away from the critical scattering length which correspond to
lifetimes of several seconds. In this parameter region, the process
investigated in this paper is, therefore, not relevant and other processes
limit the lifetime of the condensate.
However, when the scattering length is decreased the decay rate increases
significantly by several orders of magnitude and reaches up to $\Gamma \gtrsim
10^3$.
Furthermore, for blood-cell shaped BECs we identify a region of an enhanced
decay rate at $a - \acrit = 0.002 - 0.003$. This behavior is caused by the
eigenfrequencies of the oscillation modes forming the prefactor in \EQ\
\eqref{eq-Gamma-conventional}: Here, several stable eigenfrequencies of the
excited state drop down in the same parameter region. The physical
interpretation of this behavior is that the region where transitions take place
in the vicinity of the transition state becomes broader, so that the flux over
the saddle and hence the decay rate increases.

\subsubsection{The temperature regime investigated and the influence of the
strength of the DDI}

Dipolar BECs have first been realized with $^\text{52}$Cr atoms
\cite{Griesmaier2005} which possess a magnetic moment of $\mu = 6\mu_\text{B}$
($\mu_\text{B}$ is the Bohr magneton). For an exemplary chromium BEC consisting
of 10\,000 bosons, the range of the scattering length shown in
\FIG\ \ref{fig:energy-barrier} is $0.55$ Bohr radii and the energy barrier $0
\leq E^\ddagger \leq 7$ corresponds to a thermal energy $T = E^\ddagger / \kB$
of $0 \leq T \leq 141\,$nK which is about 20\% of the critical temperature ($\Tc
= 700\,$nK \cite{Griesmaier2005}).
The temperature used in \FIG\ \ref{fig:decay-rate} is $20\,$nK and the range of
the decay rate $1 \leq \Gamma \leq 3\times 10^3$ corresponds to mean lifetimes
$\tau = \Gamma^{-1}$ of $1.3\,\text{ms} \leq \tau \leq 3800\,\text{ms}$ for such
a condensate, meaning that the lifetime of the BEC is reduced to a few
milliseconds in the vicinity of the critical scattering length.

These parameters also show that the investigation of the BEC within the GPE
is justified: The frequency of the respective quasi-particle mode which is
responsible for the induced collapse shrinks to zero when the critical
scattering length is approached. Very close to this value, it is, therefore,
small and we calculate values of a few hundred to a few thousand oscillations
per unit time (in particle number scaled units) close to the critical scattering
length $\acrit$.
For the example given, these correspond to values $26.4\,\text{s}^{-1} \leq
\omega \leq 264\,\text{s}^{-1}$.
Furthermore, the temperature scale on which the thermal activation of the
quasi-particle modes has to be expected, can be estimated when one assigns
to each oscillation mode an energy of $E = \hbar  \omega$ and the corresponding
temperature $T = E/\kB$. This yields the temperature regime $0.2\,\text{nK}
\leq T = \hbar \omega/\kB \leq 2 \, \text{nK}$.
On the other hand, single-particle excitations can occur due to the thermal
energy. Although modifications are caused by the inter-particle interactions,
their contribution can be roughly estimated using the condensate fraction of a
ideal Bose gas in a harmonic trap, $N/N_0 = 1 - (T/\Tc)^3$.
The temperature regime in \FIG\ \ref{fig:decay-rate} is $0 \leq T \leq 20\,$nK,
so that the non-condensate fraction is less than $2.5 \times 10^{-5}$.
Single-particle excitations can, therefore, be neglected and the condensate can
be well described by the GPE.

Dipolar BECs are also accessible with $^\text{164}$Dy ($\mu = 10\mu_\text{B}$)
and $^{168}$Er ($\mu = 7\mu_\text{B}$) atoms \cite{Lu2011, Aikawa2012}.
Because of their larger magnetic moment and their higher masses, the
corresponding ``dipole length'' $a_\text{d}$ (see \SEC\ \ref{sec:dipolar-BEC})
is significantly larger than that of $^\text{52}$Cr. Calculating the respective
values, we obtain that it is increased by a factor of about $8.8$ for
$^\text{164}$Dy and $4.4$ for $^{168}$Er.
For these elements we, therefore, expect even larger effects.
On the one hand, this is because the range of the s-wave scattering
length in which the process presented in this paper is relevant, becomes larger
by this factor. For example the range of the scattering length in \FIGS\
\ref{fig:energy-barrier}(b) and \ref{fig:decay-rate}(b) becomes $2.40$ Bohr
radii in case of erbium and $4.78$ Bohr radii in case of dysprosium.
On the other hand, the dipolar energy scales as $E_\text{d} \sim
a_\text{d}^{-2}$. Consequently, the energy barriers discussed above are by a
factor of $1/(8.8)^{2} \approx 0.0129$ smaller for dysprosium and $1/(4.4)^{2}
\approx 0.0517$ times smaller for erbium.
Consequently, the thermal decay rates are increased significantly for these
elements.

\subsubsection{Stability threshold at finite temperatures}

The fact that a condensate at finite temperature is unstable with respect to a
coherent collapse and has a significantly shortened lifetime in the vicinity of
the critical scattering length needs to be considered when one investigates the
stability threshold of the condensate.
From a theoretical point of view this threshold is determined by a stability
analysis of the $T=0$ ground state with respect to elementary excitations.
Within numerical approaches to the GPE this can be performed using the
Bogoliubov-de Gennes equations, or in terms of variational approaches the same
is given by the eigenvalues of the Jacobian matrix of the equations of motion at
the fixed points. (In case of the variational approach, the $T=0$ stability
threshold is given by the highest-lying bifurcation of the ground state.)
In experiments, the stability threshold can be determined by successively
decreasing the s-wave scattering length to a certain value and testing whether
or not the BEC still exists there.

However, from our calculations we expect that these theoretical and
experimental methods will, in general, lead to different results due to the
finite temperature of the BEC in an experiment.
As our investigations show, the lifetime of the BEC at $T>0$ is significantly
reduced in the vicinity of the critical scattering length and it can decay in a
parameter region where the $T=0$ ground state is stable. Thus, when one
decreases the scattering length, the condensate decays because of the thermally
induced collapse before the critical value has been reached.
As a consequence, we expect that the experimentally measured values of the
stability threshold should be larger than predicted from the theory at $T=0$.
The deviation will depend on the temperature of the BEC and, for typical values
of several ten nK, it has to be expected to be on the order of a few Bohr radii.

This interpretation can be an explanation of the measurements performed by Koch
\etal\ \cite{Koch2008} who have observed such a behavior for a wide range of the
trap aspect ratio $\lambda$ in dipolar BECs. Similar observations have also been
made in BECs without long-range interaction \cite{Roberts2001}.

\section{Conclusion}

In this paper, we have investigated dipolar BECs at finite temperature within
an extended variational framework. We have shown that the excitation of
collective oscillations can induce the coherent collapse of the condensate in a
regime of the external physical parameters where the $T=0$ ground state is
stable.
Our investigations reveal that there exist -- depending on the trapping
parameters and the s-wave scattering length -- several transition states
with different rotational symmetry which can mediate collapse dynamics of the
BEC.

For BECs at finite temperature, we identified the $m=0$ collectively excited
stationary state to mediate the collapse of dipolar BECs with conventional
density distribution, and the $m=2$ state in the case of a blood-cell shaped
BEC.
The corresponding collapse dynamics shows qualitative differences: In the first
case we observe a global collapse, while in the second case the BEC collapses
locally with a d-wave symmetry.

The activation energy necessary to induce the collapse shows a significant
dependence on the trap aspect ratio of the confining trap. Depending on the
strength of the dipolar interaction its magnitude is relevant for experiments
with scattering lengths up to several Bohr radii above the critical scattering
length.
In this region the lifetime of the condensate can be significantly reduced to
the order of a ms. This means that it is experimentally demanding to reach
this regime.

Because of the gradually decreasing energy barriers the process discussed
in this paper becomes more and more important when one approaches the critical
scattering length, so that the temperature of the gas can have a direct
influence on the experimentally measured value of the stability threshold.

\begin{acknowledgments}
This work was supported by Deutsche Forschungsgemeinschaft. A.\,J.\ and M.\,K.\
are grateful for support from the Landesgraduiertenf\"orderung of the Land
Baden-W\"urttemberg.
\end{acknowledgments}

\appendix*
\section{Systematic access to excited states}\label{appendix:access}

\begin{figure}[t]
\includegraphics[width=\columnwidth]{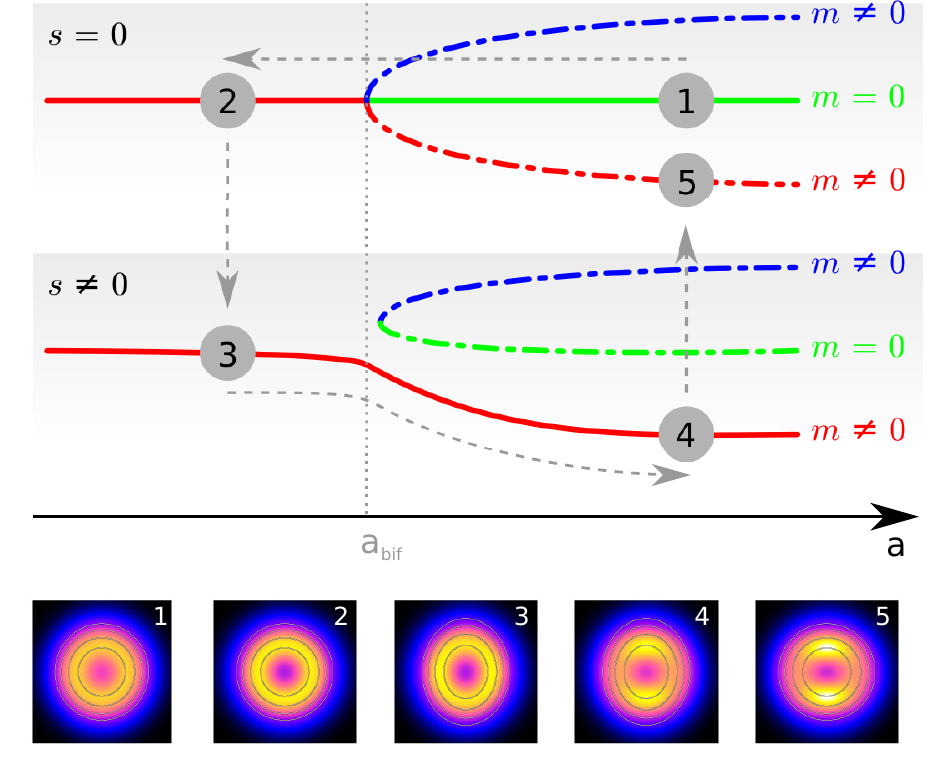}
\caption{%
(Color online) 
Schematic description of the procedure to systematically access collectively
excited stationary states which bifurcate from the ground state. Shown is
the typical bifurcation scenario between the ground state and the excited
states with $m$-fold rotational symmetry. The solid lines depict the states
which cross the bifurcation smoothly, and the dashed lines represent the
emerging states.
The numbers indicate the single steps of the procedure described in the text.
(Top) In an axisymmetric trap [$s=0$ in \EQ\ \eqref{eq:Vext}] the ground state
with $m=0$ passes the bifurcation smoothly and the $m\neq0$ states bifurcate at
a certain value $\abif$ of the s-wave scattering length.
(Center) The bifurcation scenario changes when the rotational symmetry of the
external trap is changed ($s\neq0$) in a way that the state which crosses the
bifurcation smoothly has an $m\neq0$ rotational symmetry.
(Bottom) The density profiles show the behavior of the wave function during the
steps 1--5 exemplarily for the $m=2$ bifurcation. The shape of the external trap
is indicated by the contours.
}
\label{fig:bif-scheme}
\end{figure}

Provided that the ground state of the condensate is known, there is a systematic
way to access the excited states: Therefore, we generalize a method from
Gut\"ohrlein \etal\ \cite{Gutoehrlein2013} which is based on the following
facts: On the one hand, the excited states bifurcate from the ground state at
certain values of the s-wave scattering length $a$ so that the
participating states merge at the bifurcations. One the other hand, the state
which goes through the bifurcation smoothly can be chosen by the rotational
symmetry of the external trap.

In case of an axisymmetric trapping potential [$s=0$ in \EQ\ \eqref{eq:Vext}],
the ground state of the BEC also exhibits this symmetry. Assume that this ground
state changes its stability with regard to elementary excitations with
rotational symmetry $m\neq 0$ at a certain scattering length $\abif$, then there
exist two excited states which bifurcate from the ground state in a pitchfork
bifurcation. In an axisymmetric trap, these states are physically equivalent
because they only differ by a rotation of the condensate and they can be
accessed as demonstrated in \FIG\ \ref{fig:bif-scheme}:
Starting point is the ground state of the axisymmetric trap ($s=0$) at a value
of the scattering length above $\abif$ (position 1) and we want to access the
excited state with $m\neq 0$ at the same value of the scattering length
(position 5).
In a first step, we change the scattering length to the ``other side'' of the
bifurcation (position 2) and in a second step the trap symmetry is broken by
adiabatically increasing the parameter $s$ to a sufficiently large value $s>0$
(position 3). Here, the rotational symmetry $m$ of the trap has to be chosen
according to the value which changes its stability at $\abif$ so that the ground
state naturally adopts this rotational symmetry.
In the third step, the scattering length is increased to a value above the
bifurcation (position 4) which retains the rotational symmetry and, finally, the
external trap is again changed back adiabatically to an axisymmetric shape
($s=0$).
The result at position 5 is the collectively excited state with the $m$-fold
rotational symmetry and with this procedure, we are able to access excited
states with arbitrary $m \neq 0$ bifurcating from the ground state.
In addition, the $m=0$ excited state can be found by randomly varying the
variational parameters in the vicinity of the bifurcation in a way that the
wave function keeps its axial symmetry.


\begin{thebibliography}{10}%
\makeatletter
\providecommand \@ifxundefined [1]{%
 \ifx #1\undefined \expandafter \@firstoftwo
 \else \expandafter \@secondoftwo
\fi
}%
\providecommand \@ifnum [1]{%
 \ifnum #1\expandafter \@firstoftwo
 \else \expandafter \@secondoftwo
\fi
}%
\providecommand \enquote [1]{``#1''}%
\providecommand \bibnamefont  [1]{#1}%
\providecommand \bibfnamefont [1]{#1}%
\providecommand \citenamefont [1]{#1}%
\providecommand\href[0]{\@sanitize\@href}%
\providecommand\@href[1]{\endgroup\@@startlink{#1}\endgroup\@@href}%
\providecommand\@@href[1]{#1\@@endlink}%
\providecommand \@sanitize [0]{\begingroup\catcode`\&12\catcode`\#12\relax}%
\@ifxundefined \pdfoutput {\@firstoftwo}{%
 \@ifnum{\z@=\pdfoutput}{\@firstoftwo}{\@secondoftwo}%
}{%
 \providecommand\@@startlink[1]{\leavevmode}%
 \providecommand\@@endlink[0]{}%
}{%
 \providecommand\@@startlink[1]{%
  \leavevmode
  \pdfstartlink
   attr{/Border[0 0 1 ]/H/I/C[0 1 1]}%
   user{/Subtype/Link/A<</Type/Action/S/URI/URI(#1)>>}%
  \relax
 }%
 \providecommand\@@endlink[0]{\pdfendlink}%
}%
\providecommand \url  [0]{\begingroup\@sanitize \@url }%
\providecommand \@url [1]{\endgroup\@href {#1}{\urlprefix}}%
\providecommand \urlprefix [0]{URL }%
\providecommand \Eprint[0]{\href }%
\@ifxundefined \urlstyle {%
  \providecommand \doi [1]{doi:\discretionary{}{}{}#1}%
}{%
  \providecommand \doi [0]{doi:\discretionary{}{}{}\begingroup
  \urlstyle{rm}\Url }%
}%
\providecommand \doibase [0]{http://dx.doi.org/}%
\providecommand \Doi[1]{\href{\doibase#1}}%
\providecommand \bibAnnote [3]{%
  \BibitemShut{#1}%
  \begin{quotation}\noindent
    \textsc{Key:}\ #2\\\textsc{Annotation:}\ #3%
  \end{quotation}%
}%
\providecommand \bibAnnoteFile [2]{%
  \IfFileExists{#2}{\bibAnnote {#1} {#2} {\input{#2}}}{}%
}%
\providecommand \typeout [0]{\immediate \write \m@ne }%
\providecommand \selectlanguage [0]{\@gobble}%
\providecommand \bibinfo [0]{\@secondoftwo}%
\providecommand \bibfield [0]{\@secondoftwo}%
\providecommand \translation [1]{[#1]}%
\providecommand \BibitemOpen[0]{}%
\providecommand \bibitemStop [0]{}%
\providecommand \bibitemNoStop [0]{.\EOS\space}%
\providecommand \EOS [0]{\spacefactor3000\relax}%
\providecommand \BibitemShut [1]{\csname bibitem#1\endcsname}%
\bibitem{Griesmaier2005}%
  \BibitemOpen
  \bibfield{author}{%
  \bibinfo {author} {\bibfnamefont{A.}~\bibnamefont{Griesmaier}}, \bibinfo
  {author} {\bibfnamefont{J.}~\bibnamefont{Werner}}, \bibinfo {author}
  {\bibfnamefont{S.}~\bibnamefont{Hensler}}, \bibinfo {author}
  {\bibfnamefont{J.}~\bibnamefont{Stuhler}},\ and\ \bibinfo {author}
  {\bibfnamefont{T.}~\bibnamefont{Pfau}},\ }%
  \bibfield{journal}{%
  \Doi{10.1103/PhysRevLett.94.160401}{\bibinfo {journal} {Phys. Rev. Lett.}}\
  }%
  \textbf{\bibinfo {volume} {94}},\ \bibinfo {pages} {160401} (\bibinfo {year}
  {2005})%
  \bibAnnoteFile{NoStop}{Griesmaier2005}%
\bibitem{Lu2011}%
  \BibitemOpen
  \bibfield{author}{%
  \bibinfo {author} {\bibfnamefont{M.}~\bibnamefont{Lu}}, \bibinfo {author}
  {\bibfnamefont{N.~Q.}\ \bibnamefont{Burdick}}, \bibinfo {author}
  {\bibfnamefont{S.~H.}\ \bibnamefont{Youn}},\ and\ \bibinfo {author}
  {\bibfnamefont{B.~L.}\ \bibnamefont{Lev}},\ }%
  \bibfield{journal}{%
  \Doi{10.1103/PhysRevLett.107.190401}{\bibinfo {journal} {Phys. Rev. Lett.}}\
  }%
  \textbf{\bibinfo {volume} {107}},\ \bibinfo {pages} {190401} (\bibinfo {year}
  {2011})%
  \bibAnnoteFile{NoStop}{Lu2011}%
\bibitem{Aikawa2012}%
  \BibitemOpen
  \bibfield{author}{%
  \bibinfo {author} {\bibfnamefont{K.}~\bibnamefont{Aikawa}}, \bibinfo {author}
  {\bibfnamefont{A.}~\bibnamefont{Frisch}}, \bibinfo {author}
  {\bibfnamefont{M.}~\bibnamefont{Mark}}, \bibinfo {author}
  {\bibfnamefont{S.}~\bibnamefont{Baier}}, \bibinfo {author}
  {\bibfnamefont{A.}~\bibnamefont{Rietzler}}, \bibinfo {author}
  {\bibfnamefont{R.}~\bibnamefont{Grimm}},\ and\ \bibinfo {author}
  {\bibfnamefont{F.}~\bibnamefont{Ferlaino}},\ }%
  \bibfield{journal}{%
  \Doi{10.1103/PhysRevLett.108.210401}{\bibinfo {journal} {Phys. Rev. Lett.}}\
  }%
  \textbf{\bibinfo {volume} {108}},\ \bibinfo {pages} {210401} (\bibinfo {year}
  {2012})%
  \bibAnnoteFile{NoStop}{Aikawa2012}%
\bibitem{Pedri2005}%
  \BibitemOpen
  \bibfield{author}{%
  \bibinfo {author} {\bibfnamefont{P.}~\bibnamefont{Pedri}}\ and\ \bibinfo
  {author} {\bibfnamefont{L.}~\bibnamefont{Santos}},\ }%
  \bibfield{journal}{%
  \Doi{10.1103/PhysRevLett.95.200404}{\bibinfo {journal} {Phys. Rev. Lett.}}\
  }%
  \textbf{\bibinfo {volume} {95}},\ \bibinfo {pages} {200404} (\bibinfo {year}
  {2005})%
  \bibAnnoteFile{NoStop}{Pedri2005}%
\bibitem{Nath2009}%
  \BibitemOpen
  \bibfield{author}{%
  \bibinfo {author} {\bibfnamefont{R.}~\bibnamefont{Nath}}, \bibinfo {author}
  {\bibfnamefont{P.}~\bibnamefont{Pedri}},\ and\ \bibinfo {author}
  {\bibfnamefont{L.}~\bibnamefont{Santos}},\ }%
  \bibfield{journal}{%
  \Doi{10.1103/PhysRevLett.102.050401}{\bibinfo {journal} {Phys. Rev. Lett.}}\
  }%
  \textbf{\bibinfo {volume} {102}},\ \bibinfo {pages} {050401} (\bibinfo {year}
  {2009})%
  \bibAnnoteFile{NoStop}{Nath2009}%
\bibitem{Tikhonenkov2008}%
  \BibitemOpen
  \bibfield{author}{%
  \bibinfo {author} {\bibfnamefont{I.}~\bibnamefont{Tikhonenkov}}, \bibinfo
  {author} {\bibfnamefont{B.~A.}\ \bibnamefont{Malomed}},\ and\ \bibinfo
  {author} {\bibfnamefont{A.}~\bibnamefont{Vardi}},\ }%
  \bibfield{journal}{%
  \Doi{10.1103/PhysRevLett.100.090406}{\bibinfo {journal} {Phys. Rev. Lett.}}\
  }%
  \textbf{\bibinfo {volume} {100}},\ \bibinfo {pages} {090406} (\bibinfo {year}
  {2008})%
  \bibAnnoteFile{NoStop}{Tikhonenkov2008}%
\bibitem{Dutta2007}%
  \BibitemOpen
  \bibfield{author}{%
  \bibinfo {author} {\bibfnamefont{O.}~\bibnamefont{Dutta}}\ and\ \bibinfo
  {author} {\bibfnamefont{P.}~\bibnamefont{Meystre}},\ }%
  \bibfield{journal}{%
  \Doi{10.1103/PhysRevA.75.053604}{\bibinfo {journal} {Phys. Rev. A}}\ }%
  \textbf{\bibinfo {volume} {75}},\ \bibinfo {pages} {053604} (\bibinfo {year}
  {2007})%
  \bibAnnoteFile{NoStop}{Dutta2007}%
\bibitem{Ronen2007}%
  \BibitemOpen
  \bibfield{author}{%
  \bibinfo {author} {\bibfnamefont{S.}~\bibnamefont{Ronen}}, \bibinfo {author}
  {\bibfnamefont{D.~C.~E.}\ \bibnamefont{Bortolotti}},\ and\ \bibinfo {author}
  {\bibfnamefont{J.~L.}\ \bibnamefont{Bohn}},\ }%
  \bibfield{journal}{%
  \Doi{10.1103/PhysRevLett.98.030406}{\bibinfo {journal} {Phys. Rev. Lett.}}\
  }%
  \textbf{\bibinfo {volume} {98}},\ \bibinfo {pages} {030406} (\bibinfo {year}
  {2007})%
  \bibAnnoteFile{NoStop}{Ronen2007}%
\bibitem{Goral2000}%
  \BibitemOpen
  \bibfield{author}{%
  \bibinfo {author} {\bibfnamefont{K.}~\bibnamefont{G{\'o}ral}}, \bibinfo
  {author} {\bibfnamefont{K.}~\bibnamefont{Rzazewski}},\ and\ \bibinfo {author}
  {\bibfnamefont{T.}~\bibnamefont{Pfau}},\ }%
  \bibfield{journal}{%
  \Doi{10.1103/PhysRevA.61.051601}{\bibinfo {journal} {Phys. Rev. A}}\ }%
  \textbf{\bibinfo {volume} {61}},\ \bibinfo {pages} {051601} (\bibinfo {year}
  {2000})%
  \bibAnnoteFile{NoStop}{Goral2000}%
\bibitem{Koch2008}%
  \BibitemOpen
  \bibfield{author}{%
  \bibinfo {author} {\bibnamefont{{T. Koch}}}, \bibinfo {author}
  {\bibnamefont{{T. Lahaye}}}, \bibinfo {author} {\bibnamefont{{J. Metz}}},
  \bibinfo {author} {\bibnamefont{{B. Fr{\"o}hlich}}}, \bibinfo {author}
  {\bibnamefont{{A. Griesmaier}}},\ and\ \bibinfo {author} {\bibnamefont{{T.
  Pfau}}},\ }%
  \bibfield{journal}{%
  \bibinfo {journal} {Nature Physics}\ }%
  \textbf{\bibinfo {volume} {4}},\ \bibinfo {pages} {218} (\bibinfo {year}
  {2008})%
  \bibAnnoteFile{NoStop}{Koch2008}%
\bibitem{Santos2000}%
  \BibitemOpen
  \bibfield{author}{%
  \bibinfo {author} {\bibfnamefont{L.}~\bibnamefont{Santos}}, \bibinfo {author}
  {\bibfnamefont{G.~V.}\ \bibnamefont{Shlyapnikov}}, \bibinfo {author}
  {\bibfnamefont{P.}~\bibnamefont{Zoller}},\ and\ \bibinfo {author}
  {\bibfnamefont{M.}~\bibnamefont{Lewenstein}},\ }%
  \bibfield{journal}{%
  \Doi{10.1103/PhysRevLett.85.1791}{\bibinfo {journal} {Phys. Rev. Lett.}}\ }%
  \textbf{\bibinfo {volume} {85}},\ \bibinfo {pages} {1791} (\bibinfo {year}
  {2000})%
  \bibAnnoteFile{NoStop}{Santos2000}%
\bibitem{Goral2002}%
  \BibitemOpen
  \bibfield{author}{%
  \bibinfo {author} {\bibfnamefont{K.}~\bibnamefont{G{\'o}ral}}\ and\ \bibinfo
  {author} {\bibfnamefont{L.}~\bibnamefont{Santos}},\ }%
  \bibfield{journal}{%
  \Doi{10.1103/PhysRevA.66.023613}{\bibinfo {journal} {Phys. Rev. A}}\ }%
  \textbf{\bibinfo {volume} {66}},\ \bibinfo {pages} {023613} (\bibinfo {year}
  {2002})%
  \bibAnnoteFile{NoStop}{Goral2002}%
\bibitem{Santos2003}%
  \BibitemOpen
  \bibfield{author}{%
  \bibinfo {author} {\bibfnamefont{L.}~\bibnamefont{Santos}}, \bibinfo {author}
  {\bibfnamefont{G.~V.}\ \bibnamefont{Shlyapnikov}},\ and\ \bibinfo {author}
  {\bibfnamefont{M.}~\bibnamefont{Lewenstein}},\ }%
  \bibfield{journal}{%
  \Doi{10.1103/PhysRevLett.90.250403}{\bibinfo {journal} {Phys. Rev. Lett.}}\
  }%
  \textbf{\bibinfo {volume} {90}},\ \bibinfo {pages} {250403} (\bibinfo {year}
  {2003})%
  \bibAnnoteFile{NoStop}{Santos2003}%
\bibitem{Wilson2008}%
  \BibitemOpen
  \bibfield{author}{%
  \bibinfo {author} {\bibfnamefont{R.~M.}\ \bibnamefont{Wilson}}, \bibinfo
  {author} {\bibfnamefont{S.}~\bibnamefont{Ronen}}, \bibinfo {author}
  {\bibfnamefont{J.~L.}\ \bibnamefont{Bohn}},\ and\ \bibinfo {author}
  {\bibfnamefont{H.}~\bibnamefont{Pu}},\ }%
  \bibfield{journal}{%
  \Doi{10.1103/PhysRevLett.100.245302}{\bibinfo {journal} {Phys. Rev. Lett.}}\
  }%
  \textbf{\bibinfo {volume} {100}},\ \bibinfo {pages} {245302} (\bibinfo {year}
  {2008})%
  \bibAnnoteFile{NoStop}{Wilson2008}%
\bibitem{Metz2009}%
  \BibitemOpen
  \bibfield{author}{%
  \bibinfo {author} {\bibnamefont{{J. Metz}}}, \bibinfo {author}
  {\bibnamefont{{T. Lahaye}}}, \bibinfo {author} {\bibnamefont{{B.
  Fr{\"o}hlich}}}, \bibinfo {author} {\bibnamefont{{A. Griesmaier}}}, \bibinfo
  {author} {\bibnamefont{{T. Pfau}}}, \bibinfo {author} {\bibnamefont{{H.
  Saito}}}, \bibinfo {author} {\bibnamefont{{Y. Kawaguchi}}},\ and\ \bibinfo
  {author} {\bibnamefont{{M. Ueda}}},\ }%
  \bibfield{journal}{%
  \bibinfo {journal} {New J. Phys.}\ }%
  \textbf{\bibinfo {volume} {11}},\ \bibinfo {pages} {055032} (\bibinfo {year}
  {2009})%
  \bibAnnoteFile{NoStop}{Metz2009}%
\bibitem{Lahaye2008}%
  \BibitemOpen
  \bibfield{author}{%
  \bibinfo {author} {\bibfnamefont{T.}~\bibnamefont{Lahaye}}, \bibinfo {author}
  {\bibfnamefont{J.}~\bibnamefont{Metz}}, \bibinfo {author}
  {\bibfnamefont{B.}~\bibnamefont{Fr{\"o}hlich}}, \bibinfo {author}
  {\bibfnamefont{T.}~\bibnamefont{Koch}}, \bibinfo {author}
  {\bibfnamefont{M.}~\bibnamefont{Meister}}, \bibinfo {author}
  {\bibfnamefont{A.}~\bibnamefont{Griesmaier}}, \bibinfo {author}
  {\bibfnamefont{T.}~\bibnamefont{Pfau}}, \bibinfo {author}
  {\bibfnamefont{H.}~\bibnamefont{Saito}}, \bibinfo {author}
  {\bibfnamefont{Y.}~\bibnamefont{Kawaguchi}},\ and\ \bibinfo {author}
  {\bibfnamefont{M.}~\bibnamefont{Ueda}},\ }%
  \bibfield{journal}{%
  \Doi{10.1103/PhysRevLett.101.080401}{\bibinfo {journal} {Phys. Rev. Lett.}}\
  }%
  \textbf{\bibinfo {volume} {101}},\ \bibinfo {pages} {080401} (\bibinfo {year}
  {2008})%
  \bibAnnoteFile{NoStop}{Lahaye2008}%
\bibitem{Hensler2003}%
  \BibitemOpen
  \bibfield{author}{%
  \bibinfo {author} {\bibnamefont{{S. Hensler}}}, \bibinfo {author}
  {\bibnamefont{{J. Werner}}}, \bibinfo {author} {\bibnamefont{{A.
  Griesmaier}}}, \bibinfo {author} {\bibnamefont{{P. O. Schmidt}}}, \bibinfo
  {author} {\bibnamefont{{A. G\"orlitz}}}, \bibinfo {author} {\bibnamefont{{T.
  Pfau}}}, \bibinfo {author} {\bibnamefont{{K. Rzazewski}}},\ and\ \bibinfo
  {author} {\bibnamefont{{S. Giovanazzi}}},\ }%
  \bibfield{journal}{%
  \bibinfo {journal} {Appl. Phys. B}\ }%
  \textbf{\bibinfo {volume} {77}} (\bibinfo {year} {2003})%
  \bibAnnoteFile{NoStop}{Hensler2003}%
\bibitem{Stoof1997}%
  \BibitemOpen
  \bibfield{author}{%
  \bibinfo {author} {\bibfnamefont{H.~T.~C.}\ \bibnamefont{Stoof}},\ }%
  \bibfield{journal}{%
  \bibinfo {journal} {Journal of Statistical Physics}\ }%
  \textbf{\bibinfo {volume} {87}},\ \bibinfo {pages} {1353} (\bibinfo {year}
  {1997})%
  \bibAnnoteFile{NoStop}{Stoof1997}%
\bibitem{Marquardt2012}%
  \BibitemOpen
  \bibfield{author}{%
  \bibinfo {author} {\bibfnamefont{K.}~\bibnamefont{Marquardt}}, \bibinfo
  {author} {\bibfnamefont{P.}~\bibnamefont{Wieland}}, \bibinfo {author}
  {\bibfnamefont{R.}~\bibnamefont{H\"afner}}, \bibinfo {author}
  {\bibfnamefont{H.}~\bibnamefont{Cartarius}}, \bibinfo {author}
  {\bibfnamefont{J.}~\bibnamefont{Main}},\ and\ \bibinfo {author}
  {\bibfnamefont{G.}~\bibnamefont{Wunner}},\ }%
  \bibfield{journal}{%
  \Doi{10.1103/PhysRevA.86.063629}{\bibinfo {journal} {Phys. Rev. A}}\ }%
  \textbf{\bibinfo {volume} {86}},\ \bibinfo {pages} {063629} (\bibinfo {year}
  {2012})%
  \bibAnnoteFile{NoStop}{Marquardt2012}%
\bibitem{Junginger2012d}%
  \BibitemOpen
  \bibfield{author}{%
  \bibinfo {author} {\bibfnamefont{A.}~\bibnamefont{Junginger}}, \bibinfo
  {author} {\bibfnamefont{J.}~\bibnamefont{Main}}, \bibinfo {author}
  {\bibfnamefont{G.}~\bibnamefont{Wunner}},\ and\ \bibinfo {author}
  {\bibfnamefont{T.}~\bibnamefont{Bartsch}},\ }%
  \bibfield{journal}{%
  \Doi{10.1103/PhysRevA.86.023632}{\bibinfo {journal} {Phys. Rev. A}}\ }%
  \textbf{\bibinfo {volume} {86}},\ \bibinfo {pages} {023632} (\bibinfo {year}
  {2012})%
  \bibAnnoteFile{NoStop}{Junginger2012d}%
\bibitem{Rau2010b}%
  \BibitemOpen
  \bibfield{author}{%
  \bibinfo {author} {\bibfnamefont{S.}~\bibnamefont{Rau}}, \bibinfo {author}
  {\bibfnamefont{J.}~\bibnamefont{Main}}, \bibinfo {author}
  {\bibfnamefont{H.}~\bibnamefont{Cartarius}}, \bibinfo {author}
  {\bibfnamefont{P.}~\bibnamefont{K{\"o}berle}},\ and\ \bibinfo {author}
  {\bibfnamefont{G.}~\bibnamefont{Wunner}},\ }%
  \bibfield{journal}{%
  \Doi{10.1103/PhysRevA.82.023611}{\bibinfo {journal} {Phys. Rev. A}}\ }%
  \textbf{\bibinfo {volume} {82}},\ \bibinfo {pages} {023611} (\bibinfo {year}
  {2010})%
  \bibAnnoteFile{NoStop}{Rau2010b}%
\bibitem{Kreibich2013a}%
  \BibitemOpen
  \bibfield{author}{%
  \bibinfo {author} {\bibnamefont{{M. Kreibich}}}, \bibinfo {author}
  {\bibnamefont{{J. Main}}},\ and\ \bibinfo {author} {\bibnamefont{{G.
  Wunner}}},\ }%
  \bibfield{journal}{%
  \bibinfo {journal} {J. Phys. B: At. Mol. Opt. Phys.}\ }%
  \textbf{\bibinfo {volume} {46}},\ \bibinfo {pages} {045302} (\bibinfo {year}
  {2013})%
  \bibAnnoteFile{NoStop}{Kreibich2013a}%
\bibitem{Proukakis1996}%
  \BibitemOpen
  \bibfield{author}{%
  \bibinfo {author} {\bibfnamefont{N.~P.}\ \bibnamefont{Proukakis}}\ and\
  \bibinfo {author} {\bibfnamefont{K.}~\bibnamefont{Burnett}},\ }%
  \bibfield{journal}{%
  \bibinfo {journal} {J. Res. Natl. Inst. Stand. Technol.}\ }%
  \textbf{\bibinfo {volume} {101}},\ \bibinfo {pages} {457} (\bibinfo {year}
  {1996})%
  \bibAnnoteFile{NoStop}{Proukakis1996}%
\bibitem{Griffin1996}%
  \BibitemOpen
  \bibfield{author}{%
  \bibinfo {author} {\bibfnamefont{A.}~\bibnamefont{Griffin}},\ }%
  \bibfield{journal}{%
  \Doi{10.1103/PhysRevB.53.9341}{\bibinfo {journal} {Phys. Rev. B}}\ }%
  \textbf{\bibinfo {volume} {53}},\ \bibinfo {pages} {9341} (\bibinfo {year}
  {1996})%
  \bibAnnoteFile{NoStop}{Griffin1996}%
\bibitem{Blakie2008}%
  \BibitemOpen
  \bibfield{author}{%
  \bibinfo {author} {\bibnamefont{{P. B. Blackie}}}, \bibinfo {author}
  {\bibnamefont{{A. S. Bradley}}}, \bibinfo {author} {\bibnamefont{{M. J.
  Davis}}}, \bibinfo {author} {\bibnamefont{{R. J. Ballagh}}},\ and\ \bibinfo
  {author} {\bibnamefont{{C. W. Gardiner}}},\ }%
  \bibfield{journal}{%
  \bibinfo {journal} {Adv. Phys.}\ }%
  \textbf{\bibinfo {volume} {57}},\ \bibinfo {pages} {636} (\bibinfo {year}
  {2008})%
  \bibAnnoteFile{NoStop}{Blakie2008}%
\bibitem{McLachlan1964}%
  \BibitemOpen
  \bibfield{author}{%
  \bibinfo {author} {\bibnamefont{{A. D. McLachlan}}},\ }%
  \bibfield{journal}{%
  \bibinfo {journal} {Molecular Physics}\ }%
  \textbf{\bibinfo {volume} {8}},\ \bibinfo {pages} {39} (\bibinfo {year}
  {1964})%
  \bibAnnoteFile{NoStop}{McLachlan1964}%
\bibitem{Cartarius2008a}%
  \BibitemOpen
  \bibfield{author}{%
  \bibinfo {author} {\bibnamefont{{H. Cartarius}}}, \bibinfo {author}
  {\bibnamefont{{T. Fab\v{c}i\v{c}}}}, \bibinfo {author} {\bibnamefont{{J.
  Main}}},\ and\ \bibinfo {author} {\bibnamefont{{G. Wunner}}},\ }%
  \bibfield{journal}{%
  \Doi{10.1103/PhysRevA.78.013615}{\bibinfo {journal} {Phys. Rev. A}}\ }%
  \textbf{\bibinfo {volume} {78}},\ \bibinfo {pages} {013615} (\bibinfo {year}
  {2008})%
  \bibAnnoteFile{NoStop}{Cartarius2008a}%
\bibitem{Rau2010a}%
  \BibitemOpen
  \bibfield{author}{%
  \bibinfo {author} {\bibfnamefont{S.}~\bibnamefont{Rau}}, \bibinfo {author}
  {\bibfnamefont{J.}~\bibnamefont{Main}},\ and\ \bibinfo {author}
  {\bibfnamefont{G.}~\bibnamefont{Wunner}},\ }%
  \bibfield{journal}{%
  \Doi{10.1103/PhysRevA.82.023610}{\bibinfo {journal} {Phys. Rev. A}}\ }%
  \textbf{\bibinfo {volume} {82}},\ \bibinfo {pages} {023610} (\bibinfo {year}
  {2010})%
  \bibAnnoteFile{NoStop}{Rau2010a}%
\bibitem{Junginger2012a}%
  \BibitemOpen
  \bibfield{author}{%
  \bibinfo {author} {\bibnamefont{{A. Junginger}}}, \bibinfo {author}
  {\bibnamefont{{J. Main}}}, \bibinfo {author} {\bibnamefont{{G. Wunner}}},\
  and\ \bibinfo {author} {\bibnamefont{{M. Dorwarth}}},\ }%
  \bibfield{journal}{%
  \bibinfo {journal} {J. Phys. A: Math. Theor.}\ }%
  \textbf{\bibinfo {volume} {45}},\ \bibinfo {pages} {155201} (\bibinfo {year}
  {2012})%
  \bibAnnoteFile{NoStop}{Junginger2012a}%
\bibitem{Junginger2012b}%
  \BibitemOpen
  \bibfield{author}{%
  \bibinfo {author} {\bibnamefont{{A. Junginger}}}, \bibinfo {author}
  {\bibnamefont{{M. Dorwarth}}}, \bibinfo {author} {\bibnamefont{{J. Main}}},\
  and\ \bibinfo {author} {\bibnamefont{{G. Wunner}}},\ }%
  \bibfield{journal}{%
  \bibinfo {journal} {J. Phys. A: Math. Theor.}\ }%
  \textbf{\bibinfo {volume} {45}},\ \bibinfo {pages} {155202} (\bibinfo {year}
  {2012})%
  \bibAnnoteFile{NoStop}{Junginger2012b}%
\bibitem{Haenggi1990}%
  \BibitemOpen
  \bibfield{author}{%
  \bibinfo {author} {\bibnamefont{{P. H{\"a}nggi}}}, \bibinfo {author}
  {\bibnamefont{{P. Talkner}}},\ and\ \bibinfo {author} {\bibnamefont{{M.
  Borkovec}}},\ }%
  \bibfield{journal}{%
  \bibinfo {journal} {Rev. Mod. Phys.}\ }%
  \textbf{\bibinfo {volume} {62}},\ \bibinfo {pages} {251} (\bibinfo {year}
  {1990})%
  \bibAnnoteFile{NoStop}{Haenggi1990}%
\bibitem{Wilson2009}%
  \BibitemOpen
  \bibfield{author}{%
  \bibinfo {author} {\bibfnamefont{R.~M.}\ \bibnamefont{Wilson}}, \bibinfo
  {author} {\bibfnamefont{S.}~\bibnamefont{Ronen}},\ and\ \bibinfo {author}
  {\bibfnamefont{J.~L.}\ \bibnamefont{Bohn}},\ }%
  \bibfield{journal}{%
  \Doi{10.1103/PhysRevA.80.023614}{\bibinfo {journal} {Phys. Rev. A}}\ }%
  \textbf{\bibinfo {volume} {80}},\ \bibinfo {pages} {023614} (\bibinfo {year}
  {2009})%
  \bibAnnoteFile{NoStop}{Wilson2009}%
\bibitem{Roberts2001}%
  \BibitemOpen
  \bibfield{author}{%
  \bibinfo {author} {\bibfnamefont{J.~L.}\ \bibnamefont{Roberts}}, \bibinfo
  {author} {\bibfnamefont{N.~R.}\ \bibnamefont{Claussen}}, \bibinfo {author}
  {\bibfnamefont{S.~L.}\ \bibnamefont{Cornish}}, \bibinfo {author}
  {\bibfnamefont{E.~A.}\ \bibnamefont{Donley}}, \bibinfo {author}
  {\bibfnamefont{E.~A.}\ \bibnamefont{Cornell}},\ and\ \bibinfo {author}
  {\bibfnamefont{C.~E.}\ \bibnamefont{Wieman}},\ }%
  \bibfield{journal}{%
  \Doi{10.1103/PhysRevLett.86.4211}{\bibinfo {journal} {Phys. Rev. Lett.}}\ }%
  \textbf{\bibinfo {volume} {86}},\ \bibinfo {pages} {4211} (\bibinfo {year}
  {2001})%
  \bibAnnoteFile{NoStop}{Roberts2001}%
\bibitem{Gutoehrlein2013}%
  \BibitemOpen
  \bibfield{author}{%
  \bibinfo {author} {\bibnamefont{{R. Gut\"ohrlein}}}, \bibinfo {author}
  {\bibnamefont{{J. Main}}}, \bibinfo {author} {\bibnamefont{{H. Cartarius}}},\
  and\ \bibinfo {author} {\bibnamefont{{G. Wunner}}},\ }%
  \bibfield{journal}{%
  \bibinfo {journal} {J. Phys. A}\ }%
  \textbf{\bibinfo {volume} {46}},\ \bibinfo {pages} {305001} (\bibinfo {year}
  {2013})%
  \bibAnnoteFile{NoStop}{Gutoehrlein2013}%
\end{thebibliography}
%

\end{document}